\documentclass{aa}

\usepackage{epsfig,times}

\usepackage{amssymb}

\usepackage{graphicx}

\begin{document}
\title{Convection in protoneutron stars and the structure of surface magnetic
  fields in pulsars}
\author{V.~ Urpin$^{1,2}$, J.~ Gil$^{3}$}
\offprints{V.Urpin}
\institute{$^{1}$ A.F.Ioffe Institute of Physics and Technology,
           194021 St.Petersburg, Russia \\
           $^{2}$ Departament de Fisica Aplicada, Universitat d'Alacant,
           Ap. Correus 99, 03080 Alacant, Spain \\
           $^{3}$ Institute of Astronomy, University of Zielona Gora, Lubuska 2,
           65-265 Zielona Gora}

\date{today}

\abstract{We consider generation and evolution of small-scale
magnetic fields in neutron stars. These fields can be generated by
small-scale turbulent dynamo action soon after the collapse when
the proto-neutron star is subject to convective and neutron finger
instabilities. After instabilities stop, small-scale fields should
be frozen into the crust that forms initially at high density
$\sim 10^{14}$ g/cm$^{3}$ and then spreads to the surface. Because
of high crustal conductivity, magnetic fields with the lengthscale
$\sim 1-3$ km can survive in the crust as long as 10-100 Myr and
form a sunspot-like structure at the surface of radiopulsars.
 }

\titlerunning{The structure of surface magnetic fields in pulsars}

\maketitle

\section{Introduction}

Our knowledge of the magnetic field strength in neutron stars
comes mainly from radiopulsars with measured spin-down rates. With
the assumption that the spin-down torque on the pulsar is
determined by its magnetodipole radiation (Ostriker \& Gunn 1969),
the spin period, $P$, and spin-down rate, $\dot{P}$, are related
to the field strength at the magnetic pole, $B_{d}$, by
\begin{equation}
P \dot{P} = \frac{8 \pi^{2} B_{d}^{2} R^{6}}{3 c^{2} I}
\end{equation}
where $I$ is the moment of inertia and $R$ is the radius (we
assume that the magnetic and rotation axes are perpendicular). The
magnetic fields inferred from the spin-down data range from $\sim
5 \times 10^{13}$ to $\sim 10^{8}$G but, most likely, these fields
characterize the global magnetic configuration of neutron stars
rather than a fine magnetic structure near the stellar surface.

Measurements of the spin-down rate are not, however, the only way
to obtain information about the neutron star magnetic fields.
Recent observations of the X-ray spectra features of some pulsars
provide one more opportunity to look into the magnetic field near
the surface. Absorbtion features in the spectrum of the isolated
pulsar 1E 1207.4-5209 were associated by Sanwal et al. (2002) with
atomic transitions of once-ionized helium in the neutron star
atmosphere with a strong ``surface'' magnetic field, $B_{s} \sim
1.5 \times 10^{14}$G. An estimate of the spin-down rate of this
pulsar provides the strength of the ``dipole'' magnetic field,
$B_{d} \sim (2-4) \times 10^{12}$G (Pavlov et al. 2002) that is
typical for radio pulsars of that age ($\sim 0.2-1.6$ Myr). Becker
et al. (2002) found an emission line in the X-ray spectrum of PSR
B1821-24 that could be interpreted as cyclotron emission from the
corona above the pulsar's polar cap. This emission line is likely
formed in a magnetic field $B_{s} \sim 3 \times 10^{11}$G which is
approximately two orders of magnitude stronger than the ``dipole''
field inferred from $P$ and $\dot{P}$. Haberl et al. (2003)
interpreted a broad absorption feature in the spectrum of the
isolated neutron star RBS1223 as a cyclotron line produced by
protons in the magnetic field $\sim (2-6) \times 10^{13}$ G.
These measurements provide strong evidence that the local magnetic
fields on the neutron star surface can exceed the conventional
``dipole'' field.

Another piece of evidence on the distinction between the
``dipole'' and ``surface'' magnetic fields comes from the data on
emission of radio pulsars. Recently, Gil \& Mitra (2001) and Gil
\& Melikidze (2002) argued that the formation of a vacuum gap in
radio pulsars is possible if the actual surface magnetic field
near the polar cap is very strong, $B_{s} \sim 10^{13}$G,
irrespective of the field measured from the $P$-$\dot{P}$ data.
Also, radio emission from the recently discovered pulsar PSR
J2144-3933 with the longest period 8.5 s, which lies extremely far
beyond the conventional death line, can be understood if this
pulsar has a strong surface magnetic field of complex geometry
(Gil \& Mitra 2001). Radio emission of many other pulsars which
lie in the pulsar graveyard and should be radio silent can be
explained if one adopts the model with a strong and complex
surface field with a small curvature of the field lines ($<
10^{6}$ cm). This model is consistent with the conclusion of Arons
\& Scharlemann (1979) and Arons (1993) that pulsars with very long
periods ($\geq 5$ s) require a more complex field configuration
than a dipole if pair creation is essential for the mechanism of
radio emission. Analysing the phenomenon of drifting subpulses
observed in many pulsars, Gil \& Sendyk (2000) found that their
behaviour is consistent with the vacuum gap maintained by a strong
sunspot-like magnetic field. Following this idea Gil, Melikidze \&
Mitra (2002 b) found that in the famous case of PSR 0943+10
(Deshpande \& Rankin 1999, 2001) the ``surface'' field
 should be approximately 7 times stronger than the
``dipole'' component inferred from the spin-down rate. A
sunspot-like configuration of the surface magnetic field is also
suggested by the spin-down index in some pulsars (Cheng \&
Ruderman 1993, Ruderman, Zhu \& Cheng 1998). Also, Cheng \& Zhang
(1999), analysing the X-ray emission from the polar regions of the
rotation-powered pulsars, argued that $B_{s} \sim 10^{13}$G and
the characteristic curvature of the field lines is around
$10^{5}$cm resembling the sunspot-like structure.

Rapidly growing amount of evidence on the distinction between the
local field strength at the stellar surface and the global
``dipole'' field suggests that this can be a general phenomenon in
neutron stars. In this paper, we propose a scenario of the
formation of such complex magnetic configurations with the
strength of a small scale ``surface'' field in excess of the
``dipole'' component. It is generally accepted that neutron stars
are subject to hydrodynamic instabilities soon after their birth
in the core collapse (see, e.g., Epstein 1979, Livio, Buchler \&
Colgate 1980, Burrows \& Lattimer 1986). The convective stage
lasts about 30-40 s (Miralles, Pons \& Urpin 2000, 2002) and,
under certain conditions, turbulent motions can amplify the
magnetic field via dynamo action (Thompson \& Duncan 1993, Xu \&
Busse 2001). The generated magnetic field must be frozen into the
crust that is formed in the course of neutron star cooling.
Because of high crustal conductivity, magnetic fields with a
relatively small lengthscale, $\sim 10^{5}$ cm, can survive during
the active lifetime of radio pulsars.

The paper is organized as follows. In Section 2, we consider
convection and associated dynamo action in protoneutron stars, and
estimate the field that can be generated during this stage. We
discuss the crust formation and small scale crustal magnetic
structures in Section 3. Our results are briefly summarized in
Section 4.

\section{Convection in protoneutron stars}

Neutron stars are formed in a supernova explosion associated to
the gravitational collapse of massive stars. The explosion lasts
$\sim 1$ s and is followed by the core bounce and generation of a
strong shock which heats the protoneutron star to a very high
temperature $\sim 10^{11}$ K (Burrows \& Lattimer 1986, Burrows \&
Fryxell 1992, Rampp \& Janka 2000). Hydrostatic equilibrium
settles down on a very short timescale $\sim 10^{-3}-10^{-2}$ s
but, even when the equilibrium is reached, the surface layers are
very extended, thus the radius of a protostar is $\sim 50-100$ km
instead of the canonical 10-15 km. However, compression of the
exterior zones is quite fast and, after a few seconds, the star is
compressed to the canonical radius.

Hydrodynamic instabilities in a newly born neutron star are driven
by both the lepton gradient (Epstein 1979), and the development of
negative entropy gradients, which are common in many simulations
of supernovae explosions (Bruenn \& Mezzacappa 1994, 1995, Rampp
\& Janka 2000) and evolutionary models of protoneutron stars (Keil
\& Janka 1995, Keil, Janka \& M\"{u}ller 1996, Pons et al. 1999).
Likely, both, the convective and neutron finger instabilities, can
arise in protoneutron stars (Miralles, Pons \& Urpin 2000) with
the neutron finger unstable region typically surrounding the
convective region. Initially, only the surface layer, containing
around $0.5 M_{\odot}$, is unstable but the bottom of the unstable
region spreads down to the center. Approximately in 10 s, the
whole star is subject to instability with convection, operating in the
central region of the enclosed mass $\sim 0.5 M_{\odot}$, and the
neutron finger instability dominating in the rest of the volume.
After $\sim 20$ s, the unstable region begins to shrink to the
center and, at $t\sim 40$ s, both the temperature and lepton
gradients become too smooth to maintain instabilities. Both, the
Rayleigh and Grashoff numbers are typically large in the unstable
regions, and instabilities likely do operate in a turbulent regime
(Thompson \& Duncan 1993).

During the hydrodynamic unstable phase, a protoneutron star is
opaque to neutrino, and the turbulent velocity can be estimated by
the standard mixing-length approximation (see, e.g., Schwarzschild
1958). The largest unstable lengthscale is of the order of the
pressure lengthscale, $L$, and the turbulent velocity in this
scale, $v_{L}$, can be estimated as
\begin{equation}
v_{L} \approx \frac{L}{\tau_{L}},
\end{equation}
where $\tau_{L}$ is the instability growth time (which is of the
order of the turnover time in a scale $L$). In the convectively
unstable region, we have
\begin{equation}
\frac{1}{\tau_{L}^{2}} \sim \frac{1}{\tau_{c}^{2}} = \frac{1}{3}g
\beta \frac{|\Delta \nabla T|}{T},
\end{equation}
where $\tau_{c}$ is the convection growth time, $g$ is the
gravity, $\Delta \nabla T$ is the difference between the actual
and adiabatic temperature gradients, and $\beta$ is the thermal
expansion coefficient. The turbulent velocity, $v_{L}$, varies in
time since $\Delta \nabla T$ progressively reduces due to the
neutron star cooling. Convection is a dynamical instability and
grows on a short timescale ($\sim 0.1-1$ ms when convection is
most efficient). Assuming $L \sim 1-3$ km and using the
calculations of $\tau_{c}$ (Miralles, Pons \& Urpin 2000), we can
estimate $v_{L} \sim 10^{8}$ cm/s (except the late unstable phase
when convection is almost exhausted).

In the region of a "doubly diffusive" instability often referred
to as "neutron fingers", a destabilizing influence of the lepton
number gradient usually dominates the effect of $\Delta \nabla T$.
This instability is the astrophysical analog of salt fingers that
exist in terrestrial oceans. Physically, a fluid element perturbed
downward in the proto-neutron star can thermally equilibrate more
rapidly with the background but find itself lepton-poorer and denser
and, therefore, subject to a downward force that would amplify
perturbations. This instability is typically more efficient in the
region above the convective zone, involving a larger portion of the
stellar material. With the same reasoning, we can estimate $v_{L}$
in this region substituting
\begin{equation}
\frac{1}{\tau_{L}^{2}} \sim \frac{1}{\tau_{nf}^{2}} = \frac{1}{3} g
\delta | \nabla Y|
\end{equation}
into equation (2); $\tau_{nf}$ is the neutron finger instability
growth time, $\delta$ is the chemical expansion coefficient and
$Y=(n_{e} + n_{\nu})/n$ is the lepton fraction with $n_{e}$,
$n_{\nu}$, and $n$ is the number density of electrons, neutrinos,
and baryons, respectively. The neutron finger instability is
typically slower than convection, $\tau_{nf} \sim 30-100$ ms
(except the very early and very late phases). Being slower, this
instability can nevertheless exist in a more extended region.
Using the calculations of $\tau_{nf}$ (Miralles, Pons \& Urpin
2000), we estimate  $v_{L} \sim (1-3) \times 10^{6}$ cm/s in the
neutron finger unstable region.

Likely, protoneutron stars rotate relatively rapidly (Zwerger \&
M\"{u}ller 1997, Rampp, M\"{u}ller \& Ruffert 1998) and, hence,
can be subject to the turbulent dynamo action. The initial spins
of pulsars are not well constrained by observations but, most
likely, they lie around $\sim 100$ ms (Narayan 1987). The
influence of rotation on turbulence is characterized by the Rossby
number, Ro$= P/\tau_{L}$ where $P$ is the spin period. Since there
are two unstable regions inside the proto-neutron star with
substantially different properties, the Rossby number can differ
much in the convective and neutron finger  unstable regions. In
the convective zone, we have Ro$\sim 100$, and the influence of
rotation on turbulence is probably negligible. On the contrary, in
the neutron fingers unstable region, Ro$\sim 1$ and turbulence can
be strongly modified by the Coriolis force. Therefore, the neutron
finger unstable region seems to be better suited for the
mean-field dynamo action (see Bonanno, Rezzolla \& Urpin 2003).
Note that this region is typically above the convectively unstable
region, therefore the mean-field dynamo operates mainly in the
surface layers.

In both unstable regions, however, turbulent motions can generate
turbulent magnetic fields by small-scale dynamo action. The
electrical conductivity of hot nuclear matter, $\sigma$, is
relatively high, $\sigma \approx 1.5 \times 10^{24} T_{10}^{-2}$
s$^{-1}$ where $T_{10}= T/10^{10}K$ (Baym, Pethick \& Sutherland
1971), and therefore the characteristic timescale of ohmic
dissipation is very long ($\sim 10^{5}$ yrs for the field with the
lengthscale $L \sim 1-3$ km). The magnetic diffusivity, $\eta = c^{2}/ 4 \pi
\sigma$, is small compared to viscosity, $\nu$, therefore the
dynamo operates in the regime of large magnetic Prandtl numbers,
Pr=$\nu/\eta$. There are two factors that cause viscous stresses
in protoneutron stars: neutrino transport and neutron scattering
(Thompson \& Duncan 1993). Neutrino-induced viscosity dominates on
scales larger than the neutrino mean-free path and can be very
efficient. However, even this large viscosity cannot prevent
instability on scales comparable to the pressure lengthscale
(Miralles, Pons \& Urpin 2000). The viscosity caused by neutron
scattering operates on scales shorter than the neutrino mean-free
path and is much smaller, but the magnetic Prandtl number is large
even in this case,
\begin{equation}
\mathrm{Pr}= \frac{\nu}{\eta} \approx 2.6 \times 10^{4} \rho_{14}^{5/4}
T_{10}^{-4} \gg 1,
\end{equation}
where $\rho_{14}= \rho/ 10^{14}$g/cm$^{3}$ and $\rho$ is the density; we
use the analytical fit for the coefficient of viscosity of hot nuclear matter
obtained by Cutler, Lindblom \& Splinter (1990).

There are two qualitatively different phases to a small-scale
dynamo action: the kinematic phase when the field does not provide
a noticeable influence on turbulent motions, and the dynamical
phase when velocity is affected by the Lorentz force. In the
kinematic regime, a weak seed magnetic field grows in strength
exponentially in time, while the characteristic lengthscale of a
field, $\ell_{B}$, decreases exponentially (Kazantsev 1968,
Kraichnan 1976, Kulsrud \& Anderson 1992). In protoneutron stars,
instabilities generate primary turbulent motions on scales $\sim
L$, and these motions amplify turbulent magnetic fields of the
same scale. However, the energy of such magnetic fluctuations is
transfered to the small scales after a few eddy-turnover times.
For Pr$\gg 1$, the magnetic energy is first
transfered to scales shorter than the viscous dissipative scale,
$\ell_{\nu} \sim$Re$^{-3/4}L$ (Re is the Reynolds number at the
lengthscale $L$), but greater than the magnetic dissipative
lengthscale, $\ell_{\eta} \sim \ell_{\nu}$Pr$^{-1/2}$ (see, e.g.,
Schekochihin, Boldyrev \& Kulsrud 2002). Both lengthscales,
$\ell_{\nu}$ and $\ell_{\eta}$, are very small in protoneutron
stars,
\begin{equation}
\ell_{\nu} \approx 1.9 \times 10^{-5} v_{8}^{-3/4} L_{5}^{1/4}
T_{10}^{-3/2} \rho_{14}^{15/16} \; \mathrm{cm},
\end{equation}
\begin{equation}
\ell_{\eta} \approx 1.2 \times 10^{-7} v_{8}^{-3/4} L_{5}^{1/4}
T_{10}^{1/2} \rho_{14}^{5/16} \; \mathrm{cm},
\end{equation}
where $v_{8}=v_{L}/10^{8}$cm/s, and $L_{5}= L/10^{5}$cm. Turbulent
motions with the lengthscale shorter than $\ell_{\nu}$ are
suppressed by viscosity, but magnetic fluctuations can exist if
their lengthscale is larger than $\ell_{\eta}$. Probably,
turbulence is well developed in both unstable zones, and
fluctuations spread through a wide range of scales (see, e.g.,
Thompson \& Duncan 1993). The kinematic growth phase is terminated
when the Lorentz force starts modifying the convective motions.
Since the characteristic growth time of small-scale fields, $\sim L/v_{L}$,
is much shorter than the duration of the unstable phase, the small-scale
dynamo operates likely in a nonlinear regime in both unstable regions.

In the non-linear regime, turbulence drives kinetic and magnetic
energy cascades which are quasi-steady in protoneutron stars. The
nature of these cascades in MHD turbulence has been a matter of
debate for many years. Iroshnikov (1963) and Kraichnan (1965)
suggested that in the inertial range kinetic and magnetic power
spectra are given by
\begin{equation}
E_{K}(\ell) \propto E_{M}(\ell) \propto \ell^{3/2},
\end{equation}
where $\ell$ is the lengthscale of fluctuations. On the other
hand, solar wind data suggest a spectrum $\propto \ell^{5/3}$ as
in the Kolmogorov theory. Numerical simulations (see, e.g., Kida,
Yanase \& Mizushima 1991, Haugen, Brandenburg \& Dobler 2003)
indicate that the magnetic energy spectrum probably does not show
a power law behaviour. It seems that spectra can be even shallower
than $\ell^{3/2}$ in some models. For any more or less plausible
spectra, however, the amplitude of magnetic fluctuations increases
with $\ell$ and reaches its maximum on scales comparable to the
main lengthscale of turbulence, $L$. In a saturated phase,
small-scale magnetic fields are approximately in equipartition
with velocity fluctuations,
\begin{equation}
B_{\ell} \sim B_{eq}(\ell) = \sqrt{4 \pi \rho} v_{\ell},
\end{equation}
where $B_{\ell}$ and $v_{\ell}$ are the amplitudes of magnetic and
velocity fluctuations with the lengthscale $\ell$. Equation (9)
yields only the order of magnitude estimate, since the magnetic
and kinetic energy have generally different spectra (Kida, Yanasa
\& Mizushima 1991).

Turbulence is non-stationary in both unstable zones of
protoneutron stars. It arises very rapidly soon after the
collapse, reaches some quasi-steady regime, and then goes down
when the temperature and lepton gradients are smoothed (after
$\sim 30-40$ s). The timescale required for fluid to make one turn
in a turbulent cell with the lengthscale $\ell$ can be estimated
as $\tau_{\ell} \approx \pi \ell / v_{\ell}$. This timescale
varies with time, but is typically much shorter than the
characteristic cooling timescale, $\tau_{cool}$, except the very
late phase when gradients are smoothed and instabilities are less
efficient. Therefore, turbulence can be treated in a quasi-steady
approximation during the almost whole unstable phase. Using
equation (9), we can estimate the maximum field generated in the
lengthscale $\ell \sim L$ during the quasi-steady regime as
$B_{Lc} \sim 10^{16}$ G in the convective zone, and $B_{Lnf} \sim
(1-3) \times 10^{14}$ G in the neutron finger unstable zone.
However, the temperature and lepton number gradients are
progressively reduced as the protoneutron star cools down and,
therefore, the turbulent velocity decreases as well. As the
result, the strength of small-scale magnetic fields generated by
turbulence also decreases compared to the maximum value, but
estimate (9) is still valid until the quasi-steady condition
$\tau_{cool} \gg \tau_{\ell}$, is fulfilled. We assume that this
condition breaks down at $t=t_{a}$ when $\tau_{\ell}(t)$ becomes
comparable to the cooling timescale: $\tau_{\ell}(t_{a}) \sim
\tau_{cool}(t_{a})$. Then, the turbulent velocity at $t=t_{a}$ is
given by
\begin{equation}
v_{\ell}(t_{a}) \sim \frac{\pi \ell}{\tau_{cool}(t_{a})}.
\end{equation}
We assume that the final strength of the magnetic field generated
by the small-scale dynamo, $B_{\ell}$, is of the order of
$B_{eq}(\ell)$ at $t=t_{a}$. Perhaps this estimate provides only
the upper limit on a field strength since turbulent motions can
influence magnetic fluctuations even at $t>t_{a}$. However, the
decay of turbulence is rather fast, and the kinetic energy of
turbulence rapidly becomes smaller than the magnetic energy.
Therefore, we have for the final strength of the generated
magnetic field
\begin{equation}
B_{\ell} \sim q \sqrt{4 \pi \rho} v_{\ell}(t_{a}),
\end{equation}
where $q$ is a numerical factor, $q \sim 1$. Substituting
expression (10) for $v_{\ell}(t_{a})$, we obtain
\begin{equation}
B_{\ell} \sim \frac{q \pi \sqrt{4 \pi \rho} \ell}{\tau_{cool}(t_{a})}.
\end{equation}
The final strength of the generated small-scale field turns out to
be the same for both unstable zones and decreases with decreasing
lengthscale. For the largest turbulent scale, $\ell =L \sim 1-3$
km, estimate (12) yields $B_{L} \sim 3 \times 10^{13} q$ G if
$\tau_{cool}$ is of the order of a few seconds. Note that our
estimate is in contrast to that of Thompson \& Duncan (1993) who
assumed that the strength of the generated field is approximately
given by its maximum value.

\section{Crust formation and crustal small-scale
magnetic fields}

It is difficult to predict final disposition of the magnetic field
after the turbulent motions inside the protoneutron star stop.
Most likely, there exists a wide spectrum of magnetic fluctuations
with the lengthscale shorter than $L$ and with the field strength
given by equation (12). When convection is exhausted, these
small-scale fluctuations evolve under the influence of the ohmic
dissipation and buoyancy. After a magnetohydrodynamic
quasi-equilibrium is established, one expects that magnetic loops
densely fill the volume and surface of the star, although the
field strength may vary considerably, depending on the degree of
intermittency of the generated small-scale field. In liquid
nuclear matter, the ohmic decay timescale of fluctuations with the
lengthscale $\ell$ is
\begin{equation}
\tau_{\rm O}(\ell) \sim \frac{4 \pi \sigma \ell^{2}}{c^{2}} =
2.8 \times 10^{7} T_{10}^{-2} \left( \frac{\ell}{L} \right)^{2}
{\rm yrs},
\end{equation}
and dissipation is important only for fluctuations with very
small $\ell$.

Likely, the crust formation provides the most important influence
on the evolution of small-scale magnetic fields at this stage.
Approximately at the age $\sim 40-50$ s (soon after convection
stops), the neutron star cools down to the internal temperature
$\sim (1-3) \times 10^{10}$K (Pons et al. 1999). At such
temperature, neutrons and protons can form nuclei and clusters in
the nuclear matter with the density $\sim 10^{14}$ g/cm$^{3}$.
When the neutron star cools down to a lower temperature, nuclei
can be formed at $\rho < 10^{14}$ g/cm$^{3}$ as well. Note that
nuclear composition at high density depends generally on the
pre-history of the neutron star. For instance, the composition of
"ground state" matter (Negele \& Vautherin 1973) differs
noticeably from that of "accreted" matter (Haensel \& Zdunik
1990). The Coulomb interaction of nuclei leads to the crystal
formation even at a relatively high temperature. Crystallization
occurs when the ion coupling parameter $\Gamma = Z^{2} e^{2}/(a
k_{B} T)$ reaches the critical value $\Gamma = \Gamma_{m} \approx
170$ (Slattery, Doolen \& De Witt 1980); $a= (3/4 \pi
n_{i})^{1/3}$ is the mean inter-ion distance, $n_{i}$ and $Z$ are
the number density and charge number of ions, respectively; $T$ is
the temperature, and $k_{B}$ is the Boltzmann constant. Then, the
crystallization temperature is
\begin{equation}
T_{m} = \frac{Z^{2} e^{2}}{a k_{B} \Gamma_{m}} \approx
1.3 \times 10^{7} Z^{5/3} \mu_{e}^{-1/3} \rho_{12}^{1/3}
\frac{170}{\Gamma_{m}} \;\; K,
\end{equation}
where $\mu_{e}$ is the number of baryons per one electron, and
$\rho_{12} = \rho /10^{12}$ g/cm$^{3}$. For the density $\rho \sim
10^{13}-10^{14}$ g/cm$^{3}$, the crystallization temperature is of
the order of $10^{10}$ K (Baiko \& Yakovlev 1996). Therefore, the
crust formation starts almost immediately after the end of the
convective phase, and the magnetic fields generated by convective
motions should be frozen into the crust. Solidification proceeds
rather rapidly, and the outer boundary of the crust reaches the
density $\sim 10^{10}$ g/cm$^{3}$ after $\sim 1$ day in the case
of standard cooling.

In crystal layers, the magnetic field evolution is mainly
determined by ohmic dissipation. The crust electric conductivity
$\sigma_{c}$, can be expressed in terms of the electron
relaxation time, $\tau_{e}$ (see, e.g., Baiko \& Yakovlev 1996),
\begin{equation}
\sigma_{c} \approx 1.5 \times 10^{22} x^{2} \beta \left(
\frac{\tau_{e}}{10^{-16} {\rm s}} \right) \; {\rm s}^{-1},
\end{equation}
where $x= p_{F}/m_{e} c$, $\beta = v_{F}/c$; $p_{F}$ and $v_{F}$
are the electron Fermi momentum and velocity, respectively. In a
high density region, electrons are ultrarelativistic and $\beta
\approx 1$. Using expression (15), we can estimate the ohmic decay
timescale of magnetic fluctuations with the lengthscale $\ell$ in
the crust,
\begin{equation}
\tau_{\rm Oc}(\ell) = \frac{4 \pi \sigma_{c} \ell^{2}}{c^{2}}
\approx 0.7 \times 10^{5} x^{2} \beta \left( \frac{\tau_{e}}{10^{-16}
{\rm s}} \right) \ell_{5}^{2} \;\; {\rm yrs},
\end{equation}
where $\ell_{5} = \ell/10^{5}$cm. This timescale is short
for short lengthscale fluctuations, but $\tau_{\rm Oc }$ can be
much longer than the cooling timescale for fluctuations with
relatively large $\ell$. For instance, the decay of fluctuations
with $\ell \sim 10^{3}-10^{5}$ cm proceeds certainly on a timescale
longer than the time required for the formation of a well developed
crust. Therefore, turbulent magnetic fields with relatively large
lengthscales are likely frozen into the crystallized matter soon
after convection stops.

Further evolution of turbulent magnetic fields deposed in the
neutron star crust depends on conductive properties of the crustal
matter and the cooling scenario. The behaviour of such fields is
governed by the standard induction equation where only the ohmic
dissipation is included and is qualitatively similar to the
behaviour of a large scale crustal magnetic field considered in
detail by Urpin \& Konenkov (1997). We refer the results of this
paper in what follows. We consider the evolution of turbulent
fields for the neutron star model with the standard cooling since
this cooling scenario can better account for the available
observational data. Note that models with accelerated cooling
always lead to a slower decay of the magnetic field (Urpin \& Van
Riper 1993) and, as a result the small-scale magnetic structure
can survive for a longer time in such models. The crustal
conductivity is determined by scattering of electrons on phonons
and impurities. Scattering on phonons dominates the conductivity
while the neutron star is relatively hot whereas impurities give
the main contribution at a low crustal temperature (see, e.g.,
Yakovlev \& Urpin 1980). The effect of impurities on the
conductivity is characterized by the impurity parameter $Q$,
\begin{equation}
Q= \frac{1}{n} \sum_{n'} n' (Z-Z')^{2},
\end{equation}
where $n'$ is the number density of an interloperspecies of charge
$Z'$, $n$ and $Z$ are the number density and charge of the dominant
ion species; summation is over all species.  Most likely, $Q$ ranges
from 0.001 to 0.1 within the crust.
\begin{figure}
\includegraphics[width=9.5cm]{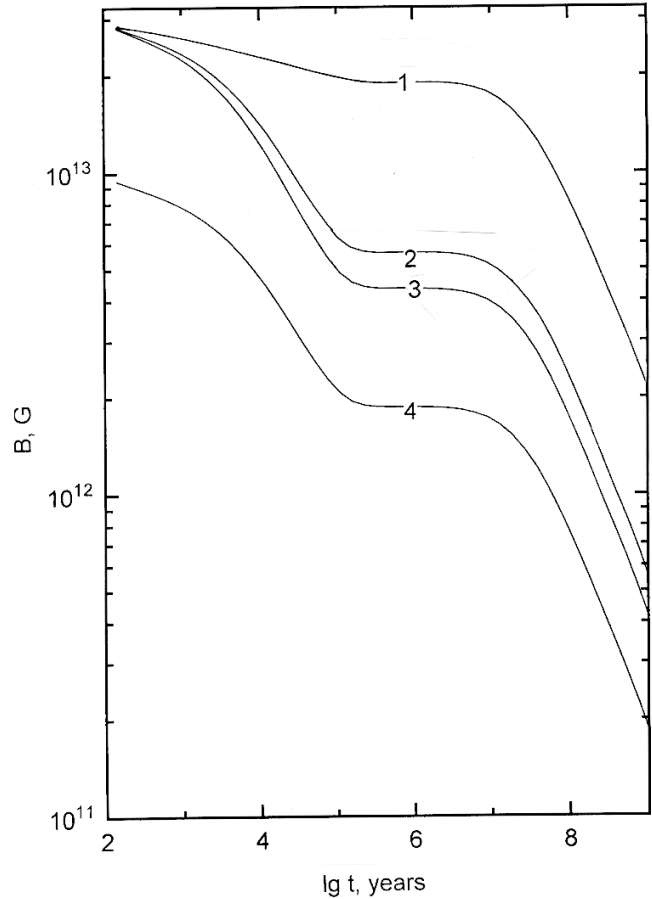}
\caption{The evolution of the surface magnetic field strength $B$ for
different initial lengthscales $L$. The
decay curves are shown for $L=2$ km (curve 1), $1.2$ km (curves 2 and
4), and $1.0$ km (curve 3). The initial field is $3 \times 10^{13}$ G
for the curves 1, 2 and 3, and $10^{13}$ G for the curve 4. The crustal
impurity parameter is $Q=0.01$.}
\end{figure}

In Fig.1, we plot the time dependence of the surface strength of a
small-scale magnetic field for a neutron star model with mass $M=1.4
M_{[\odot]}$ and with the equation of state of Pandharipande and Smith
(see, e.g., Pandharipande, Pines \& Smith 1976). The radius of this model
is $R=15.98$ km. For the sake of simplicity, we assume in calculations
that the dependence of a field on polar and azimuthal coordinates is
sinusoidal with the wavelength equal to the main scale of turbulence, $L$,
and use the so called local approximation in these two directions. This
is a sufficiently good approximation since the radius of the star is
substantially larger than $L$ and the thickness of the crust. We also
assume that the initial radial depth of a small-scale field is equal to
the main scale of turbulence, $L$, and varies within the range 1-2 km.
This range of depth corresponds to the density from $\approx 5
\times 10^{12}$ g/cm$^{3}$ to $\approx 10^{14}$ g/cm$^{3}$ for the
PS-model. The initial radial dependence of the field is chosen in
accordance with the model proposed by Urpin \& Konenkov (1997).
Note that main conclusions of our paper are not sensitive to a
particular choice of this dependence but are flexible to the value
of $L$.

The magnetic field evolution is shown for three initial depths
of the field, $L=1$ km (this depth corresponds to the
density $\rho \approx 5 \times 10^{12}$ g/cm$^{3}$ in the crust),
$L=1.2$ km ($\rho \approx 10^{13}$ g/cm$^{3}$), and $L= 2$km
($\rho \approx 10^{14}$ km). For the purpose of illustration,
the decay in the case $L=1.2$ km is shown for two initial field
strength, $3 \times 10^{13}$ G and $10^{13}$ G (curves 2 and 4,
respectively). During the initial stage ($t \leq
10^{5}$ yrs), the crustal conductivity is determined by scattering
of electrons on phonons and is relatively low for all calculated
models. Therefore, the field decay can be essential during
this stage, and the surface field strength can weaken by a factor
$\approx 2-7$, depending on the lengthscale of the initial
turbulent field. Obviously, a decrease during the initiale stage
is smaller for the field with larger $L$. For instance, a
small-scale field with $L=1$ km is reduced by a factor $\approx 7$
after $10^{5}$ yrs whereas the field with $L=2$ km decreases
less than twice after the same time. Note that for $L > 2$ km
the decrease during the initial stage is practically negligible,
and the field at $t \sim 10^{5}$ yrs is only a bit lower than the
initial field.

After $\sim 10^{5}$ yrs, the dominant conductivity mechanism
changes from electron-phonon to electron-impurity scattering. As a
result, the conductivity increases and the rate of field decay
slows down. We calculate the field decay during this stage using
an intermediate value of the impurity parameter, $Q= 0.01$ that
characterizes not a very poluted crust. For larger $Q$, the decay
is more rapid (see, for comparison, Urpin \& Konenkov 1997). The
evolution of small-scale fields shows the presence of flat
segments of the decay curves at $t \sim 0.1-100$ Myr like those of
a large-scale field. The length of a plateau depends on the
impurity parameter, $Q$. The lower that $Q$ gets, the longer the
plateau on the corresponding decay curve is. During the impurity
dominating stage, the decay turns out to be extremely slow, and
the characteristic decay time can be as long as 10-100 Myr.

These simple model calculations show very clearly that turbulent
magnetic fields with the lengthscale of the order of the
turbulence main lenghtscale in proto-neutron stars ($\sim 1-3$ km)
can survive in the crust during a very long time $\sim 10-100$ Myr
that is generally comparable to the active lifetime of
radiopulsars. The field strength in such magnetic spots on the
surface can reach $5 \times 10^{12}-2 \times 10^{13}$ G depending
on the radius of a spot and can be larger than (or comparable to)
the strength of the dipole field.

\section{Conclusion}

We have considered the formation and evolution of small-scale
magnetic structures in the neutron star surface layers. A wide
spectrum of these structures can be generated by the small-scale
turbulent dynamo action during the unstable phase that lasts $\sim
30-40$ s after the neutron star birth. There are two substantially
different unstable regions in the proto-neutron star, with the
convective instability active in the inner region and the
neutron-finger instability more efficient in the outer region.
Generally, the small-scale dynamo can generate small-scale
magnetic structures in both unstable zones. Due to high
conductivity of nuclear matter, the lengthscale of generated
magnetic structures spreads from the main scale of turbulence, $L
\sim 1-3$ km (comparable to the pressure scale height), to
extremely short lengthscales determined by ohmic dissipation.
After instabilities stop, structures with short lengthscales decay
on a short timescale $\propto \ell^{2}$ due to finite electrical
conductivity whereas fields with larger lengthscales can survive
for a longer time. The crust formation that starts almost
immediately after instabilities stop may have the decisive
influence on the evolution of such magnetic fields. These magnetic
structures can be frozen into the crystallized matter and then
evolve in the crustal layers. The neutron star cooling increases
the conductivity of the crust and, as a result, small-scale
magnetic structures decay extremely slowly. Our calculations show
that structures with the lengthscale $L \sim 1-3$ km can survive
as long as 10-100 Myr that is basically comparable to the active
life-time of radiopulsars.

The strength of the magnetic field generated by the small-scale
dynamo action is approximately determined by equipartition and
decreases with the decreasing lengthscale. For magnetic structures
with $L \sim 1-3$ km, the magnetic field can be as strong as $\sim
5 \times 10^{12}-2 \times 10^{13}$ G even for radiopulsars as old
as $\sim 10-100$ Myr. Note that the decay of small-scale magnetic
fields is qualitatively similar to that of the large-scale crustal
field considered by Urpin \& Konenkov (1997). The origin of a
large-scale field in neutron stars is still debatable, but it is
possible that this field was generated by some mechanism in the
layer with $\rho < 2 \times 10^{14}$ g/cm$^{3}$ that corresponds
to the crust (see, e.g., Bonanno, Rezzolla \& Urpin 2003). Then, the
thickness of a layer occupied by the large-scale field can generally
differ from $L$. If the initial depth of a large-scale field is
smaller than $L$, then the small-scale field decreases
slower, and we observe a radiopulsar with magnetic "spots" where
the field is stronger than the dipole field inferred from the
spin-down rate. On the contrary, if the initial depth of a large-scale
field is larger than $L$, then small-scale
structures decay faster than the dipole field, and the resulting
magnetic field becomes more regular with the age.

Small-scale magnetic structures with $L \sim 1-3$ km and $B \sim 5
\times 10^{12}-2 \times 10^{13}$ G at the surface may have an
important influence on many properties of radiopulsars. As already
mentioned in the Introduction, this range of the surface magnetic
field is favourable for the inner vacuum gap formation in pulsars
(Gil \& Melikidze 2002). The sparking discharge of this gap
produces filaments of electron-positron plasma, whose presence
seems absolutely necessary for generation of coherent pulsar radio
emission. Moreover, the ${\bf E} \times {\bf B}$ drift of spark
plasma should be manifested as the observed subpulse drift,
provided that the actual surface field has some degree of axial
symmetry, with a tendency to converge at the local pole  (see Gil,
Melikidze \& Geppert, 2003 for review). These authors assumed that
one small-scale structure (a "spot") coincides to some extent with
the canonical (dipolar) polar cap in the sence that the magnetic field
lines form a complex magnetic configuration near the pole but
connect smoothly with a subset of open dipolar field lines at a
larger radius (Gil, Melikidze \& Mitra 2002a). This means that the
actual polar cap is defined by those non-dipolar field lines which
penetrate the light cylinder.

A study of small-scale magnetic structures can also provide
information regarding the very early evolutionary stage of neutron
stars since the observed fields were frozen into the crust soon
after the neutron star birth. For example, the presence of
small-scale magnetic structures in radiopulsars can be a good
evidence that proto-neutron stars pass the turbulent stage in
their evolution.

\section*{Acknowledgement}

One of the authors (V.U.) thanks the University of Alicante for
hospitality and the Spanish Ministry of Science and Technology for
a financial support (grant AYA2001-3490-C02-02). This paper is
supported in parts by the Polish KBN Grant 2 P03D 008 19 and the
grant 04-02-16243 of the Russian Foundation of Basic Research.
J.G. acknowledges the renewal of the Alexander von Humboldt fellowship.

{}

\end{document}